\documentclass[a4paper,11pt]{article}
\usepackage{aaskaiid,subfigure,orcidlink}
\title{Cm-wavelength Studies of Molecular Gas and Star Formation at High Redshift with the SKA}
\ShortTitle{Molecular gas at high-z with SKA}

\author[1,2]{Ran Wang\orcidlink{0000-0003-4956-5742}}
\ShortName{Wang, R. et al.} 
\author[3]{Jeff Wagg}
\author[4]{Mark T. Sargent\orcidlink{0000-0003-1033-9684}}
\author[5]{Manuel Aravena}
\author[6]{Mamta Pandey-Pommier\orcidlink{0000-0001-5829-1099}}
\author[7]{Elisabete da Cunha\orcidlink{0000-0001-9759-4797}}
\author[8]{Eric Murphy\orcidlink{0000-0001-7089-7325}}
\author[9]{Dominik Riechers\orcidlink{0000-0001-9585-1462}}
\author[10]{Jacqueline Hodge}

\affiliation[1]{Department of Astronomy, School of Physics, Peking University, Beijing 100871, China}
\emailAdd{rwangkiaa@pku.edu.cn}
\affiliation[2]{Kavli Institute of Astronomy and Astrophysics at Peking University,
No.5 Yiheyuan Road, Haidian District, Beijing, 100871, China}
\affiliation[3]{Laboratoire Lagrange, OCA, CNRS, Blvd de l'Observatoire, CS 34229, F-06304 Nice cedex 4, France}
\affiliation[4]{Institute of Physics, Laboratory of Astrophysics, \'Ecole Polytechnique F\'ed\'erale de Lausanne (EPFL), Observatoire de Sauverny, Versoix CH-1290, Switzerland}
\affiliation[5]{Instituto de Estudios Astrofísicos, Facultad de Ingeniería y Ciencias, Universidad Diego Portales, Av. Ejército 441, Santiago, 8370191, Chile}
\affiliation[6]{Pole Scientific, University Catholic of Lyon- University of Lyon, 10 place des Archives 69288, Lyon, France}
\affiliation[7]{International Centre for Radio Astronomy Research, University of Western Australia, 35 Stirling Hwy, Crawley, WA 6009, Australia}
\affiliation[8]{National Radio Astronomy Observatory, 520 Edgemont Road, Charlottesville, VA 22903, USA}
\affiliation[9]{Institut f\"{u}r Astrophysik, Universit\"{a}t zu K\"{o}ln, Z\"{u}lpicher Stra\ss e 77, D-50937 K\"{o}ln, Germany}
\affiliation[10]{Leiden Observatory Gorlaeus Building room BW.3.34 Einsteinweg 55, 2333 CC Leiden, The Netherlands}

\abstract{The Square Kilometre Array will be a revolutionary instrument for the study of gas in the distant 
Universe. At frequencies below $\sim$50 GHz, observations of redshifted emission from low-{\it J} transitions 
of CO, HCN, HCO+, and HNC, etc. provide insight into the kinematics and mass budget of the cold, 
dense star-forming gas in galaxies. Over the past decade, sensitive imaging using ALMA has detected 
and resolved the redshifted high-{\it J} molecular CO line emission and far-infrared fine structure 
lines in samples of galaxies over a wide redshift range, shedding light on active star-formation 
processes at the early epoch of galaxy evolution. In recent years, increasing numbers of young galaxies at 
high redshift are discovered by JWST, which significantly improved our knowledge of different galaxy 
populations across cosmic time. In this updated chapter of the SKA science book, we would 
like to highlight the importance of studies of the low-{\it J} molecular lines in high-z galaxies 
using SKA toward high frequencies, discussing the request of frequency coverage beyond 15 GHz 
and emphasizing its crucial role in exploring the cold molecular gas content in the young galaxy populations in the early universe and investigating the regions of active-star 
formation using molecular CO and various dense gas tracers. }

\begin{document}
\maketitle
\newcommand{\actaa}{Acta Astron.} 
\newcommand{\araa}{ARA\&A} 
\newcommand{\aar}{A\&ARv} 
\newcommand{\aapr}{A\&ARv} 
\newcommand{\ab}{Astrobiol.} 
\newcommand{\aj}{AJ} 
\newcommand{\apj}{ApJ} 
\newcommand{\apjl}{ApJL} 
\newcommand{\apjs}{ApJSS} 
\newcommand{\ao}{Appl. Opt.} 
\newcommand{\apss}{Astro. \& Space Sci.} 
\newcommand{\aap}{A\&A} 
\newcommand{\aaps}{A\&AS.} 
\newcommand{\baas}{Bull. Am. Astron. Soc.} 
\newcommand{\caa}{Chinese A\&A} 
\newcommand{\cjaa}{Chinese J. A\&A} 
\newcommand{\cqg}{Class. Quantum Gravity} 
\newcommand{\gal}{Galaxies} 
\newcommand{\gca}{Geo. Cosmo. Acta} 
\newcommand{\icarus}{Icarus} 
\newcommand{\jcap}{JCAP} 
\newcommand{\jgr}{J. Geophys. Res.} 
\newcommand{\jgrp}{J. Geophys. Res. Planets} 
\newcommand{\jqsrt}{J. Quant. Spectrosc. Radiat. Transf.} 
\newcommand{\memsai}{Mem. SAIt} 
\newcommand{\mnras}{MNRAS} 
\newcommand{\nat}{Nature} 
\newcommand{\nastro}{Nat. Astron.} 
\newcommand{\ncomms}{Nat. Commun.} 
\newcommand{\nphys}{Nat. Phys.} 
\newcommand{\na}{New Astron.} 
\newcommand{\nar}{New Astron. Rev.} 
\newcommand{\physrep}{Phys. Rep.} 
\newcommand{\pra}{Phys. Rev. A} 
\newcommand{\prb}{Phys. Rev. B} 
\newcommand{\prc}{Phys. Rev. C} 
\newcommand{\prd}{Phys. Rev. D} 
\newcommand{\pre}{Phys. Rev. E} 
\newcommand{\prx}{Phys. Rev. X} 
\newcommand{\prl}{Phys. Rev. Let.} 
\newcommand{\psj}{Planet. Sci. J.} 
\newcommand{\planss}{Planet. Space Sci.} 
\newcommand{\pnas}{Proc. Natl Acad. Sci. USA} 
\newcommand{\procspie}{Proc. SPIE} 
\newcommand{\pasa}{PASA} 
\newcommand{\pasj}{PASJ} 
\newcommand{\pasp}{PASP} 
\newcommand{\rmxaa}{RMXAA} 
\newcommand{\sci}{Science} 
\newcommand{\sciadv}{Sci. Adv.} 
\newcommand{\solphys}{Sol. Phys.} 
\newcommand{\sovast}{Soviet Ast.} 
\newcommand{\ssr}{Space Sci. Rev.} 
\newcommand{\uni}{Universe} 

\section{Introduction:star formation and galaxy evolution at the early epoch}
Measurements of star formation and the cold gas content are key elements for understanding mass assembly
and galaxy formation in the early universe. Sensitive submillimeter/millimeter surveys provide the most
efficient way to detect the infrared thermal dust emission and obscured star formation in high
redshift galaxies \citep[e.g., ][]{hodge20,bouwens22,herrera25}. 
The millimeter bands of the modern interferometer arrays, such as
the Northern Extended Millimeter Array (NOEMA) and the Atacama Large Millimeter/submm Array (ALMA), and
the cm bands on the Karl G. Jansky Very Large Array (VLA) allow detection of CO transitions from bright millimeter sources, which trace the molecular gas in the young galaxies in the early
universe \citep{decarli16,decarli22,pavesi18,riechers19,birkin21,urquhart22}. 
These studies, in combination with the deep optical/near-infrared observations, 
reveal high molecular gas fractions in star-forming galaxies in early cosmic epochs \citep[e.g.,][]{tacconi20}. 

The large collecting area and broad frequency coverage of ALMA allow observations of 
the high-z galaxies with star formation rates (SFR) down to a few to tens $\rm M_{\odot}\,yr^{-1}$ 
through the bright [CII] 158$\mu$m fine structure line and dust continuum, which detected and resolved the star 
formation activity in the main sequence of star forming galaxies over a wide 
redshift range \citep{walter16,aravena16,schreiber17,herrera25}, measured the star formation rate in the earliest 
Lyman-break galaxies (LBGs), Ly$\alpha$ emitters, and quasar hosts \citep{maiolino15,smit18,hygate23,wang24,fujimoto24}, and 
provided the first constraints on the number density of the obscured star-forming systems 
beyond z$\sim$6 \citep{uzgil21,barrufet23}.  
These observations of the past decade significantly improved our knowledge of the
efficiency and distribution of star formation in early galaxies. 
The compact and intense star formation discovered through ALMA fine structure line and
continuum observations reveals the formation of massive bulges in the early evolutionary phases of these luminous
objects \citep{lelli21,hodge25,bodansky25}. 
The kinematics of the star-forming interstellar medium (ISM) traced by [CII] and high-{\it J} CO lines have provided the first measurements of galaxy dynamical masses and offered insight into star formation and active galactic nuclei (AGN) feedback in high-$z$ galaxies \citep{pensabene20,jones21,neeleman21,walter22,salak24,spilker25}.  

The new generation of optical and infrared telescopes, such as the James Webb Space Telescope (JWST) and Euclid satellite, plus
the upcoming ground-based facilities, including the Vera C. Rubin observatory and the
Extremely Large Telescope (ELT), enable deep observations of the unobscured star
formation (via UV continuum, Ly$\alpha$, and H$\alpha$ lines, etc.) and stellar light from
the young galaxies in the early universe. 
Over the past few years, the new JWST infrared data has revolutionized our 
understanding of galaxy evolution. It discovered the first light from the earliest star-forming systems and
supermassive black holes, reported a dominant fraction of disk galaxies at z$>$3 in 
systems with stellar masses of $\rm M_{*}\geq 10^{9}\,M_{\odot}$ \citep{ferreira23}, and
probed the main sequence of star formation galaxies at z$\sim$12 \citep{conselice25}. 
These deep infrared observations provide key constraints on the evolution of galaxy morphology and distributions of stellar mass over cosmic time,  
by enabling detailed measurements of their effective radii and S\'ersic indices and tracing the evolution of galaxies from disk-like to bulge-dominated systems \citep{ormerod24,martorano25}. While extensive studies have been carried out at optical and infrared wavelengths, radio studies focused on the molecular gas in high-redshift galaxies have been limited, lacking a comprehensive understanding of the cold gas reservoirs that fuel star formation in radio galaxies.

The high frequency bands of the Square Kilometre 
Array (Band 5b with SKA1-MID and Band 6 in a future array upgrade, Figure 1) with improved $\mu$Jy-level sensitivity 
compared to the existing facilities (e.g., the Green Bank Telescope, VLA) will open a unique opportunity for observations of 
the ground state transitions of the important molecules, such as CO, HCN, and HCO+ etc., 
in young galaxies at high redshift. While the obscured star formation and the high-{\it J} molecular 
lines are imaged with ALMA, the stellar components are well studied with JWST, and the HI gas is expected to 
be investigated with the low frequency bands of SKA, observations 
of the low-{\it J} transitions of these molecules will directly constrain the mass budget of the molecular gas, and together with the submm/mm line and dust continuum measurements, will provide diagnostics to determine the physical conditions and excitation mechanisms 
of the star-forming ISM. Together, these multi-wavelength observations will fill key gaps in our understanding of early galaxy formation and the regulation of star formation across cosmic time.

\section{Molecular CO line emission with the SKA}

The Band 5b receiver installed on SKA1-MID will cover the frequency range of 8.3 to 15.4 GHz, which 
will allow surveys of the CO {\it J\rm =1-0} line at z$\geq$6.5. In this redshift range, the molecular gas is 
mainly explored through the CO {\it J \rm=6-5} and {\it J\rm =7-6} transitions which fall in the sensitive millimeter 
windows of ALMA and NOEMA \citep[e.g.,][]{strandet17,decarli22,ono22,hashimoto23}. 
The CO {\it J\rm =1-0} or {\it J\rm =2-1} lines were searched in several systems including quasar host galaxies 
and star-forming Ly$\alpha$ emitters \citep[LAEs,][]{wagg09,wagg12,zavala22,kaasinen24} with the GBT or VLA, and detected  
only in the most luminous object with a infrared luminosity of $\rm L_{IR}\sim10^{13}\,L_{\odot}$ 
and warm dust temperature of $\rm T_{dust}=61\,K$ \citep[J2348$-$3054,][]{kaasinen24}.
Observations of the CO {\it J\rm =1-0} line with SKA1-MID in the more common galaxy population with 
$\rm L_{IR}< 10^{12}\,L_{\odot}$ or SFR about or less then tens $\rm M_{\odot}\,yr^{-1}$ at this highest redshift 
range remain difficult. As was discussed in \citet[][AASKA14-161]{wagg15}, even with a very 
long integration time of hundreds of hours, a low kinetic temperature of $\rm T_{kin}\leq20-30K$ will result 
in undetectable line flux against the cosmic microwave background \citep[CMB,][]{dacunha13}. 
Therefore, the best targets for SKA1-MID to conduct a CO {\it J\rm =1-0} line survey are still the 
most luminous starburst systems at z$\sim$6.5-7 with SFRs of a few hundreds to a thousand $\rm M_{\odot}\,yr^{-1}$ and with 
a warm kinetic temperature of $\rm T_{kin}\geq 40K$. 
Such objects are rare at the earliest epoch and likely to be formed in the highly biased regions 
within $\rm 800\,Myr$ after the Big Bang \citep{springel05,riechers13,barrufet23,wang24}. 
However, every such example is of great importance 
to probe the formation of the first massive galaxies. If other parameters such as the stellar mass, 
galaxy dynamical mass, dust mass, and metallicity could be determined with deep ALMA and 
JWST observations, the CO {\it J\rm =1-0} line detection will add a key constraint for detailed studies 
of the gas fraction, gas-to-dust ratio, star formation efficiency, gas excitation and the $\rm CO-to-H_{2}$ 
conversion factor ($\rm \alpha_{CO}$), providing crucial insights into the physical conditions of the starburst galaxies and their molecular gas contents. 

\begin{figure}[htbp]
    \centering
	\includegraphics[width=0.7\columnwidth]{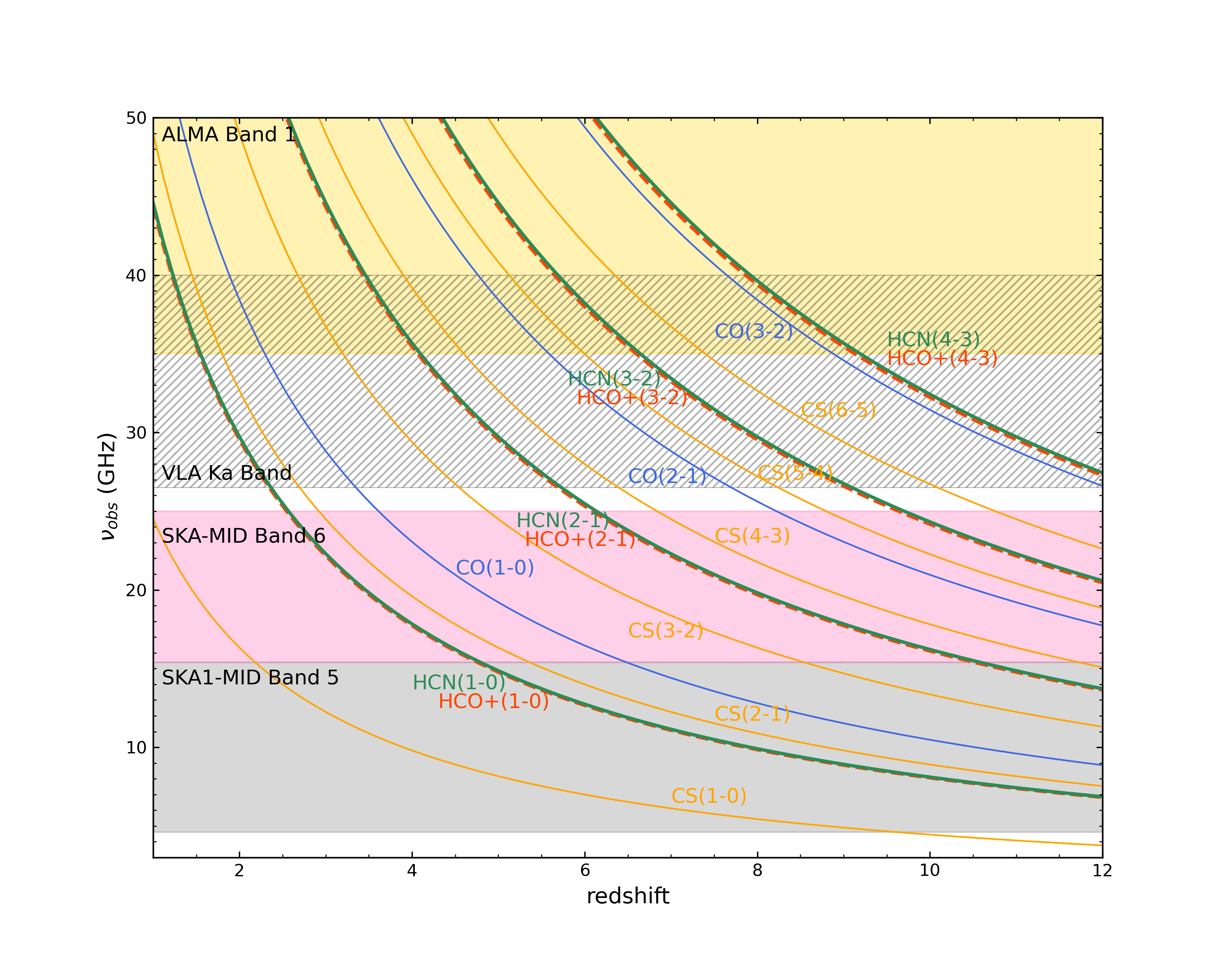}
    \caption{Observing frequencies of the low-{\it J} transitions of CO, HCN, HCO$^{+}$, and CS vs. 
    redshift (Updated from Wagg et al. 2015). The grey region shows the Band 5a and 5b  coverage of 
    4.6 to 15.4 GHz in SKA1-Mid. The pink region denotes the frequency range of 
    15.4 to 25 GHz proposed for Band 6 of future SKA. The hatched area represents the frequency range 
    of VLA Ka band receiver (26.5 to 40 GHz), and the yellow region denotes the coverage of 35 to 52 GHz for ALMA band 1.}
    \label{fig:band}
\end{figure}

With the availability of the Band 6 receivers (15.4 to 25 GHz) in the future, 
it would become possible to expand CO {\it J\rm =1-0} line surveys to samples of dusty star-forming galaxies at redshift down to z$\sim$3.6. 
For the warm star forming systems, the observed CO line flux will not be 
significantly diminished by the CMB effect in this redshift range (Figure 2).
These, together with deep line and continuum measurements in optical to millimeter bands, 
will allow research programs to address several key aspects in galaxy evolution, as discussed ahead.

Firstly, the CO {\it J\rm =1-0} line observation in a wide redshift range will allow a systematic study of the  
evolution of gas fraction, gas-to-dust ratio, and star formation efficiency in the most active star-forming populations 
over cosmic time. Secondly, the CO {\it J\rm =1-0} line, in combination with the the mid and high-{\it J} CO transitions and fine 
structure lines (e.g., [CII], [CI], [OI] etc.) detected with ALMA or NOEMA, will 
provide a full diagnostic of the star-forming ISM \citep{carilli13,yang17,boogaard20,li24,hagimoto25}. These can be compared to the 
predictions from different models to determine the local 
conditions (e.g., density, kinetic temperature, radiation field, etc.) of the ISM and to 
investigate the roles of different gas heating and excitation mechanisms, 
such as photodissociation regions (PDR) created by UV photons, shocks powered 
by feedback from supernova or active galactic nuclei (AGN), or X-ray dominated region 
created by AGN \citep{meijerink07,meijerink13,pound08,pound23}.

These studies will be particularly important for the population of AGN host galaxies 
discovered at high redshift. Active star formation was detected in the host galaxies of 
a large fraction of the optically selected quasars through thermal dust continuum 
and the [CII] line \citep[e.g.,][]{decarli18,wang24}. Molecular gas traced by mid-J CO transitions in these systems suggests a high 
star formation efficiency and quick gas depletion \citep{bischetti21,bertola24}. However, as the mid and high-{\it J} CO lines 
are mainly associated with dense gas, observations of the ground state CO transition will provide 
a more reliable measurement of the total molecular gas mass, addressing the question of whether AGN hosts tend to have lower gas 
fractions compared to normal galaxies at similar redshifts and with similar stellar masses. 

Lastly, for the most luminous objects with SFR of hundreds to thousands $\rm M_{\odot}\,yr^{-1}$,  
we will expect to resolve the CO {\it J\rm =1-0} line at $\rm 0.1''\sim0.2''$ resolution with a reasonable 
signal-to-noise ratio (e.g., SNR$>$3). The resolved line image and velocity field will 
allow a detailed study of gas dynamics, including measurements of rotation curves 
and comparisons of gas kinematics traced by different CO transitions and fine structure lines. 
Moreover, the CO {\it J\rm =1-0} line maps, together with the sensitive submm/mm images at similar angular resolution, 
will provide maps of line ratios, star formation efficiency, as well as a resolved diagnostic of the gas conditions, 
including gradients of gas density and temperature etc. This will further address how the local ISM 
conditions change with the intensity of the star formation activity, galaxy structure, and gas kinematics. 
The high-redshift luminous objects detected from existing/future submillimeter/millimeter single-dish surveys provide the primary 
targets for such studies. E.g., large-area surverys carried out using the South Pole Telescope \citep[SPT, ][]{everett20} and 
Atacama Cosmology Telescope \citep[ACT, ][]{naess25} discovered the brightest high-redshift dusty galaxies in the southern hemisphere, which are  
usually gravitationally lensed and hosting intense star formation with a warm dust temperature \citep[e.g.,][]{marsden14, reuter20}.

Detection of the CO {\it J\rm =1-0} line from a star-forming galaxy with SFR on orders 
of 10 $\rm M_{\odot}\,yr^{-1}$ at z$\sim$4 to 6 will require between hundreds to a thousand hours of exposure time with Band 6 with the SKA1 AA4 configuration, and will still be limited to the warm objects with $\rm T_{kin}\geq 40\,K$ considering the CMB effect (Figure 2). 
However, considering the increasing of star formation efficiency and decreasing of gas metallicity toward high redshift which could result in a high radiation field intensity, such warm temperature may be more common among the star-forming galaxies in the early universe \citep{bethermin15,jarugula21}.
As deep exposure is required to detect CO {\it J\rm =1-0} from such population, the observation may be first considered for the fields with known overdensity of star-forming galaxies, 
e.g., the regions with overdensity of Lyman break galaxies  
identified from JWST or other deep optical/near-IR surveys, overdensities of submm/mm sources, or the nearby fields around luminous quasars 
or massive starburst systems that are supposed to be formed in the most massive halos at the early epoch. 
This will make efficient use of the telescope time and 
field of view (e.g., $\rm \sim 4'$ at 20 GHz). Deep CO {\it J\rm =1-0} line observations with the SKA will detect the 
gas-rich obscured/unobscured star-forming systems in such regions, as well as providing very deep measurements 
of the radio continuum with the maximal instantaneous bandwidth of 5 GHz. 
If the frequency coverage could be expanded to 50 GHz, the CO {\it J\rm =1-0} line 
could be widely searched in the large population of star-forming galaxies on the 
main sequence at z$\sim$2 where the CMB effect is negligible.  
We will expect a more efficient high-z CO {\it J\rm =1-0} survey in the phase of SKA2 where the sensitivity 
will be improved by a factor of ten. 

The ground-transition of molecular CO also provide a window to search for molecular 
gas outflows in the young galaxies at high redshift which directly trace the feedback from both 
AGN activity and intense star formation \citep{veilleux20}. 
High velocity components were detected in the mid-J CO 
and [CII] 158$\mu$m line spectra of a number of high-redshift quasars which suggest molecular and neutral gas outflows on galactic scale and powered 
by the luminous AGN \citep[e.g., ][]{brusa18,li21,bischetti25}. 
This provides the first evidence that AGN regulate the surrounding gas reservoir in the early phase of black hole and galaxy evolution.
The molecular outflows from the high-z dusty star forming galaxies are mainly detected through absorption lines of OH and $\rm H_{2}O$ molecules\citep{spilker18,jones19,nianias24}.  
Neutral gas outflow in the star-forming galaxies at z$>$4 was suggested by the broad component in 
the stacking [CII] spectrum \citep{gallerani18,ginolfi20} and was detected in the [CII] emission from the z=5.5 main-sequence 
galaxy HZ4 \citep{herrera21}. 
Moreover, extended CO or [CII] structures on $\geq$10 kpc scales were also reported in several quasar and star-forming galaxies at high redshift, which may 
indicate gas injection to the circumgalactic medium by previous outflows \citep{emonts16,emonts19,fujimoto19,ginolfi20,li21,herrera21,bischetti24}. 
The CO {\it J\rm =1-0} transition will offer a direct probe of the cold, low-excitation gas component, which potentially carries the bulk of the outflowing mass and remains largely unexplored in the high-redshift galaxies. These will provide key complement to quantify the impact of AGN and star formation feedback on galaxy evolution.

\begin{figure}[htbp]
    \centering
	\includegraphics[width=0.7\columnwidth]{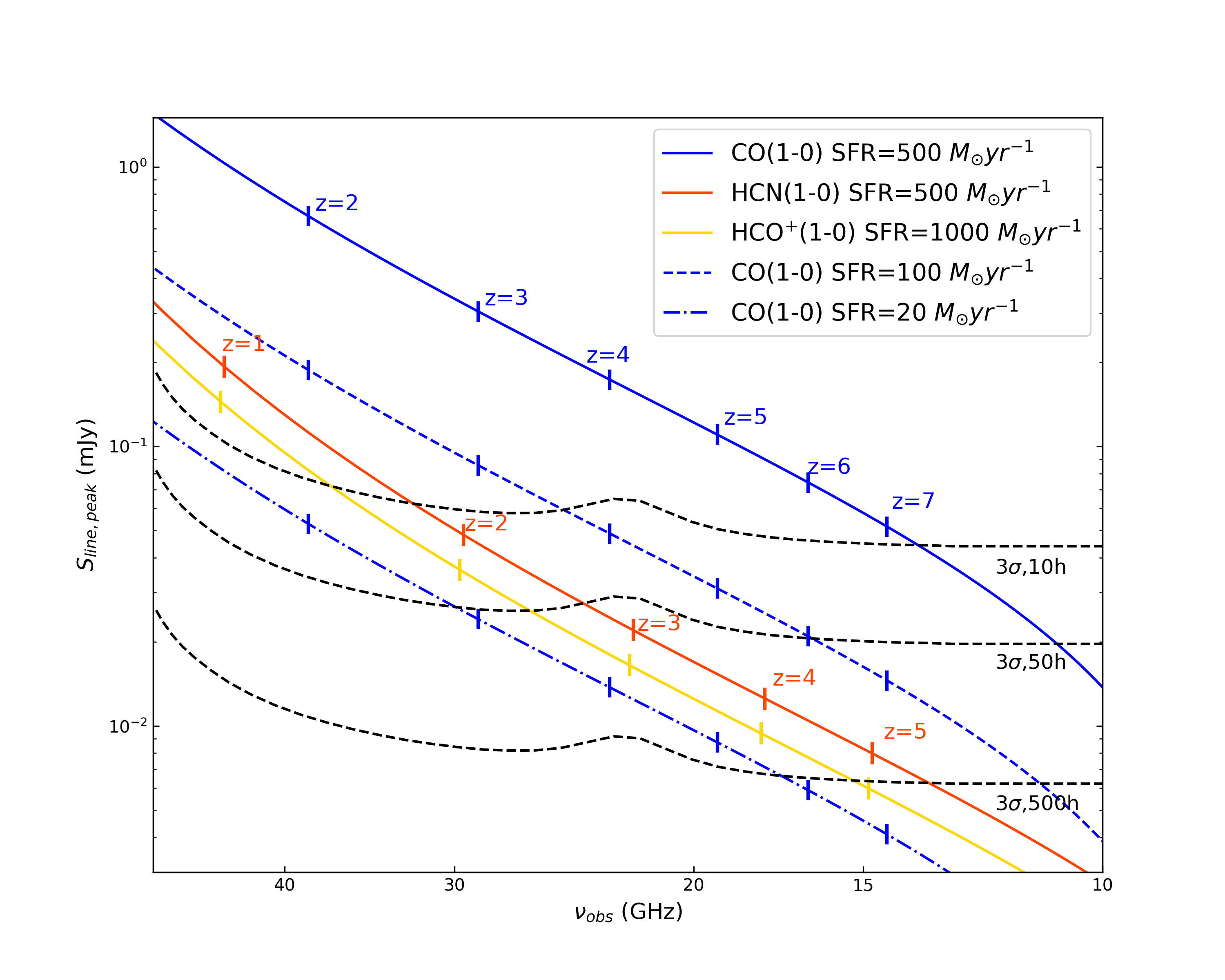}
    \caption{The observed peak line flux densities of CO {\it J\rm =1-0}, HCN {\it J\rm =1-0}, and HCO$^{+}$ {\it J\rm =1-0} against CMB 
    at different observing frequencies/redshift. The blue solid, dashed, and dash-dotted lines represent CO {\it J\rm =1-0} transition from 
    galaxies with SFRs of 500 $\rm M_{\odot}\,yr^{-1}$, 100 $\rm M_{\odot}\,yr^{-1}$, and 20 $\rm M_{\odot}\,yr^{-1}$, respectively. 
    The blue vertical bars mark the redshifts of z=2 to 7. The orange and yellow lines shows the 
    peak flux densities of HCN {\it J\rm =1-0} from a galaxy with SFR=500 $\rm M_{\odot}\,yr^{-1}$ and HCO$^{+}$ {\it J\rm =1-0} 
    with SFR=1000 $\rm M_{\odot}\,yr^{-1}$, respectively. The orange and yellow vertical bars marks the redshifts of z=1 to 5. We corrected for CMB effect adopting a 
    kinetic temperature of $\rm T_{kin}=40 K$ \citep{dacunha13}. The black dashed lines 
    show the 3$\sigma$ detection limits with on-source integration time of 10, 50, and 500 hours. These are 
    estimated from the rms line sensitivity in SKA Memo 20-01.
    }
    \label{fig:flux}
\end{figure}

\section{Dense star-forming gas}

In addition to CO transitions, the relation between star formation and molecular gas is also widely explored using  
dense gas tracers, such as HCN, HCO$^+$, HNC, and CS etc. \citep{gao04,wu10,zhang14,rybak22,neumann25,rybak25}. 
The line luminosities of the low-{\it J} rotational transitions of these molecules 
with high critical densities above $\rm 10^{4}\,cm^{-3}$ are 
linearly correlated with star formation rate \citep{gao04,garcia12,kamenetzky16}. 
Therefore, they can directly constrain the total mass of the dense molecular gas 
and star formation efficiency in actively star-forming regions. 
However, as the line luminosities of these molecules are much fainter than those of the CO molecule, detections are 
still limited to the sample of nearby star-forming galaxies. Observations of the low-{\it J} transitions of 
HCN, HCO$^{+}$, and HNC lines were carried out only on a number of 
objects at z$>$2 which are intrinsically luminous in infrared thermal 
dust emission with high SFRs of a few hundreds to thousands $\rm M_{\odot}\,yr^{-1}$ 
and/or are gravitationally lensed \citep{carilli05,solomon05,rybak22,rybak25}. Rybak et al. (2022) reported VLA 
observations of the HCN {\it J\rm =1-0}, HCO$^{+}$ {\it J\rm =1-0}, and HNC {\it J\rm =1-0} lines from six lensed dusty star-forming galaxies with only two detections, which suggests a low dense gas fraction. A recent study including more objects with detections of higher-J transitions of these tracers further reveal a wide range of dense gas fractions \citep{rybak25}. These measurements also indicate a shorter dense-gas depletion time scale in these high-z dusty star-forming systems compared to that of nearby galaxies.

The Band 5b and future Band 6 receivers of SKA will cover the bright dense 
gas tracers such as HCN {\it J\rm =1-0} and HCO$^{+}$ {\it J\rm =1-0} at z$\geq$2.5. However, 
in the phase of SKA1, a detection of the HCN {\it J\rm =1-0} line at z$\sim$4-5 in systems with 
SFR$\rm \sim 500\,M_{\odot}\,yr^{-1}$ will in practice require hundreds of hours of integration time. 
Therefore, a search for these dense molecular lines from the brightest objects could be carried out synergistically with the deep, small-area continuum surveys which will also provide deep measurements of the free-free SFRs from distant galaxies\citep[e.g.,][]{Algera01.2026.SKA}.
With the improved sensitivity compared to that of the VLA at similar frequencies, we will expect a higher 
detection rate to increase the HCN{\it J\rm =1-0} and HCO+ {\it J\rm =1-0} sample size in this redshift range. 
This will significantly improve the measurements of dense gas fractions and better constrain 
the luminosity relationships between the IR dust continuum and these dense gas tracers 
for ultraluminous dusty star-forming galaxies at high redshift. 
More efficient searches of these two lines as well as 
the ground state transitions of other fainter species (e.g., CS {\it J\rm =1-0}, HNC {\it J\rm =1-0}) should be carried out with SKA2, 
to explore the dense molecular gas content and star formation efficiency in 
the population of normal star-forming galaxies in a wider redshift range. 

\section{Predictions for SKA spectral line surveys} 

In order to predict the detectability of molecular lines in future SKA observations, 
we first estimate the peak flux densities of the CO {\it J\rm =1-0} line emission in star forming 
galaxies with SFR of $\rm 500\,M_{\odot}\,yr^{-1}$, $\rm 100\,M_{\odot}\,yr^{-1}$, and $\rm 20\,M_{\odot}\,yr^{-1}$ from z$<$2 to z$\geq$7, 
as well as HCN {\it J\rm =1-0} in galaxies with $\rm SFR=500\,M_{\odot}\,yr^{-1}$ and HCO$^{+}$ {\it J\rm =1-0} with $\rm SFR=1000\,M_{\odot}\,yr^{-1}$ from z$\leq$1 to z$\geq$7. These SFRs can be converted to $\rm 8\mu m-1000\mu m$ infrared luminosity ($\rm L_{IR}$) adopting a Chabrier Initial Mass Function \citep[i.e., $\rm SFR/M_{\odot}\,yr^{-1}\approx1.09\times10^{-10} L_{IR}/L_{\odot}$][]{chabrier03}, and represent extreme starbursts, ultraluminous infrared galaxies (ULIRGs, $\rm L_{IR}\geq 10^{12}\,L_{\odot}$), 
and more typical infrared galaxies (LIRGs, $\rm L_{IR}\geq 10^{11}\,L_{\odot}$), respectively. 
We calculate the CO {\it J\rm =1-0}, HCN {\it J\rm =1-0}, and HCO$^{+}$ {\it J\rm =1-0} line luminosities in $\rm K\,km\,s^{-1}\,pc^{2}$ adopting the literature 
relationships between line and IR luminosities \citep{gao04,garcia12,kamenetzky16,neumann25}, and the peak 
line flux densities are derived assuming a linewidth of $\rm FWHM=300\,km\,s^{-1}$. 
Finally, we calculate the observed line flux densities 
against CMB at different redshift, following the formula described in Da Cunha et al. (2013). 
As discussed in Wagg et al. (2015), it is difficult to detect the line against CMB in the systems with low 
kinetic temperature at high redshift. Here we adopt $\rm T_{kin}=40\,K$ and develop the discussion 
focussing only on objects with a warm kinetic temperature for the observations using Band 5b and Band 6.
This high kinetic temperature is likely to be a reasonable assumption for the high-z dusty star-forming  
systems and quasar host galaxies with intense star formation \citep[e.g.,][]{zavala22}.
A $\Lambda -$dominated cosmological 
model with $\rm \Omega_{\Lambda}=0.7$, $\rm \Omega_{M}=0.3$, and $\rm H_{0}=70\,km\,s^{-1}\,Mpc^{-1}$ is adopted here. 
We plot the peak flux densities of CO {\it J\rm =1-0}, HCN {\it J\rm =1-0}, and HCO$^{+}$ {\it J\rm =1-0} and the corresponding observing frequencies at different 
redshifts in Figure 2. We also add the 3$\sigma$ detection limits for on-source integration time of 10 hours, 50 hours, and 500 
hours. The 1$\sigma$ rms sensitivity for SKA between 10 GHz and 50 GHz are taken from SKA Memo 20-01, which assumed 133 SKA-Mid dishes, 
1 hour on-source integration, and a channel width of 100 $\rm km\,s^{-1}$. 

Figure 2 demonstrates that, with the high-frequency end of Band 5b on SKA1-MID, it will be possible to detect CO {\it J\rm =1-0} line emission from 
most IR luminous starburst systems at z$\sim$7 within 10 hours of integration time. 
E.g., with the empirical relations described above, an SFR of $\rm 500\,M_{\odot}\,yr^{-1}$ corresponds to IR luminosity of $\rm L_{IR}\simeq4.6\times10^{12}\,L_{\odot}$ and a CO {\it J\rm =1-0} luminosity of $\rm L'_{CO{\it J\rm =1-0}}\simeq4.6\times10^{10}\,K\,km\,s^{-1}\,pc^{2}$\citep{kamenetzky16}. 
Such systems at z$\sim$7 will have a CO {\it J\rm =1-0} line peak flux density of $\rm S_{line,peak}\sim51\mu Jy$ over the line emitting channels.
For a more precise calculation using the online SKA sensitivity calculator, in 10 hours of 
on-source integration time, we can expect a 1$\sigma$ rms of 11 $\rm \mu Jy\,beam^{-1}$ per 100 $\rm km\,s^{-1}$, assuming 
the AA4 configuration with 133 15-m antennas and pwv$=$10 mm. 
Considering a FWHM CO {\it J\rm =1-0} source size of $\rm \lesssim 0.5''$ (the typical source 
size from ALMA [CII] observations of z$\sim$7 galaxies, Fudamoto et al. 2022; Wang et al. 2024), 
we apply Briggs weighting with robust$=$1, and tapering to 
an angular resolution of $\rm 1.2''\times 1.1''$ in the calculation. This will allow us to detect the 
line at $\rm \sim 4.5\sigma$ in three consecutive 100 $\rm km\,s^{-1}$ channels. 
In case the Band 5b receiver will also be installed on the 64 MeerKAT antennas, we will expect to reduce the integration time to 5.8 hours to reach the above sensitivity with the whole 197-antenna array \citep[assuming an array System Equivalent Flux Density of 2.77 Jy,][]{pellegrini21}.

\begin{figure}[htbp]
    
    \includegraphics[width=0.45\columnwidth]{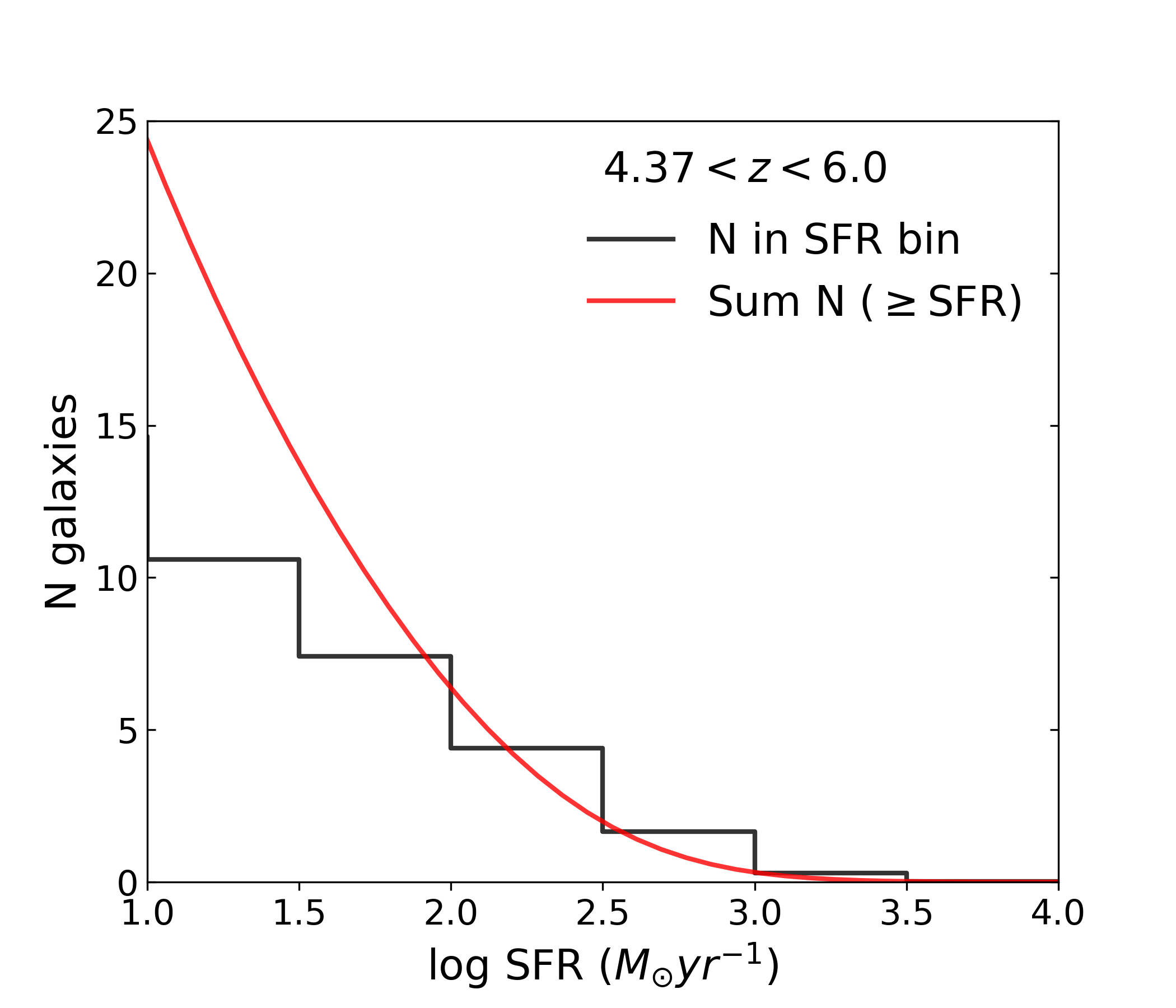}
    \includegraphics[width=0.45\columnwidth]{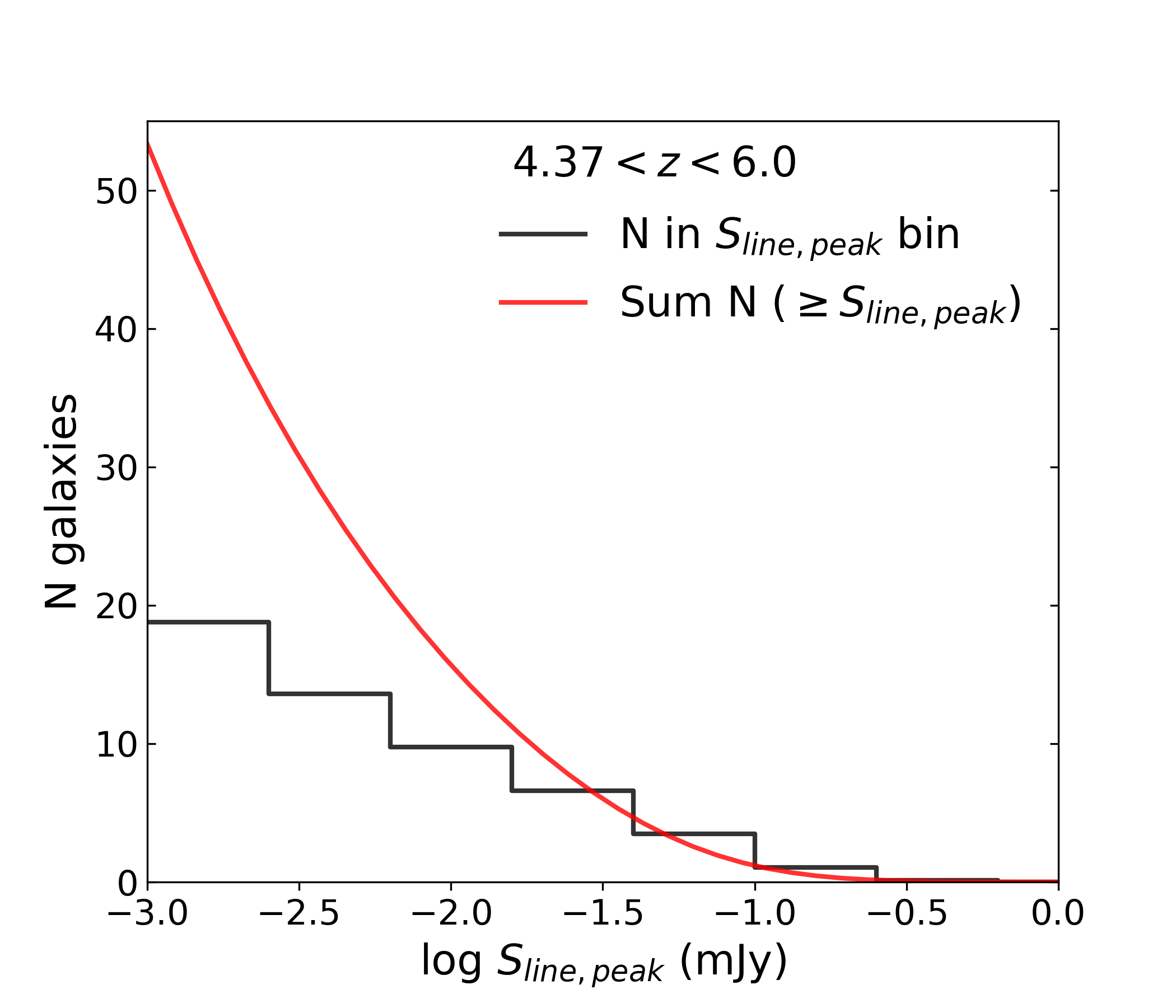}
    \caption{Number of galaxies that are expected to be covered in a single pointing field of view for Band 6 observation with SKA at different SFRs (left) and CO {\it J\rm =1-0} line peak flux densities ($\rm S_{line,peak}$, right). The redshift range of z=4.37 to 6 corresponds to the frequency coverage of the CO {\it J\rm =1-0} line in a bandwidth of 5 GHz. In the two panels, the black histograms show the number of sources in each SFR or $\rm S_{line,peak}$ bins, and the red lines show the accumulated numbers integrated above the SFR or $\rm S_{line,peak}$ limits. These are calculated based on the IR luminosity function from the ALPINE survey\citep{gruppioni20}, and the SFRs and $\rm S_{line,peak}$ are converted to $\rm L_{IR}$ using the empirical relations from the literature \citep[][see the text for details]{chabrier03,kamenetzky16}. }
\end{figure}

In a CO {\it J\rm =1-0} line survey using Band 6, we should be able to detect the line from the population of 
galaxies with $\rm SFR\geq 100\,M_{\odot}\,yr^{-1}$ at z$\sim 3.6-6$ in tens of hours of integration time. 
In Figure 3, we estimate the numbers of galaxies that are expected to be covered in a single pointing field of view at different SFRs and CO {\it J\rm =1-0} line peak flux densities. Here we assume a bandwidth of 5GHz in one frequency setup centered at 19GHz covering the redshift range of 4.37$<$z$<$6.0. The SFRs and the CO {\it J\rm =1-0} line peak flux densities are converted to IR luminosities using a Chabrier IMF and the CO {\it J\rm =1-0} to IR luminosity relation from \citet{kamenetzky16}, respectively. A factor of 1.3 is adopted here to convert the literature $\rm 42.5\mu m -122.5\mu m$ far-infrared luminosity to $\rm L_{IR}$\citep{garcia12}. Then source numbers are integrated using the IR luminosity function from the ALPINE survey \citep{gruppioni20}. According to these estimates, an object with $\rm SFR\geq100\,M_{\odot}\,yr^{-1}$ will have CO {\it J\rm =1-0} line peak flux density of $\rm S_{line,peak}\geq0.03\,mJy$ at z$=$5. This could be detected at $\rm \geq 3\sigma$ in $\sim$30 hours of integration time in AA4 configuration. We will expect to cover about 6 such galaxies in a single pointing of random field in this redshift range. 

For the most luminous objects, e.g. with $\rm SFR\geq 500\,M_{\odot}\,yr^{-1}$ at z$\sim$5, 
in AA4 configuration and 50 hours of integration time, we will expect to well resolve the CO {\it J\rm =1-0} line emission and measure the line surface brightness 
at $\rm \gtrsim 3.5\sigma$ with an rms of 7.6 $\rm \mu Jy\,beam^{-1}$ per 100 $\rm km\,s^{-1}$ channel. 
Here we adopt a tapering beam size of $\rm 0.2''\times 0.2''$, and assume that we can resolve the CO {\it J\rm =1-0} source into 4 beams. 

As described in Sections 2 and 3, deep integrations of hundreds to a thousand hours in Band 6 will allow us to detect the CO {\it J\rm =1-0} line 
from the star forming galaxies on the main sequence at z$\sim$3.6-6 with SFRs on order of 10 $\rm M_{\odot}\,yr^{-1}$, or HCN {\it J\rm =1-0} line from 
objects with SFR $\geq$ a few hundreds $\rm M_{\odot}\,yr^{-1}$ at 2.5$<$z$\leq$5. 
Based on the prediction from SKA Memo 20-01, we will expect an rms line sensitivity of 2.4 $\rm \mu Jy\,beam^{-1}$ per 100 $\rm km\,s^{-1}$ channel at 19.2 GHz in 500 hours of integration time. Assuming a typical FWHM line 
width of 300 $\rm km\,s^{-1}$, the 3$\rm \sigma$ upper limit corresponds to a CO {\it J\rm =1-0} line luminosity 
of $\rm 3.0\times10^{9}\,K\,km\,s^{-1}\,pc^{2}$ at z$\sim$5, or HCN {\it J\rm =1-0} luminosity 
of $\rm 2.7\times10^{9}\,K\,km\,s^{-1}\,pc^{2}$ at z$\sim$3.6 (adopting $\rm T_{kin}=40\,K$ for CMB correction). 
According to Figure 3, we will expect to detect tens of CO {\it J\rm =1-0} sources at this sensitivity level in one tuning at 4.37$<$z$<$6. 

\section{Summary}
In this chapter, we have evaluated the capabilities of the SKA to study key molecular lines, such as CO{\it J\rm =1-0}, HCN{\it J\rm =1-0}, and HCO$^{+}${\it J\rm =1-0} in high-redshift star-forming galaxies using SKA1-MID Band 5b and the future Band 6. Our analysis shows that SKA will enable the detection of CO{\it J\rm =1-0} emission in luminous starburst galaxies out to $z \sim 7$, and in more typical star-forming galaxies (with SFR $\gtrsim$ 100 M$_\odot$ yr$^{-1}$) at intermediate redshifts ($z \sim 3.6$–6) using Band 6. Deep integrations will further allow the detection of dense gas tracers, such as HCN{\it J\rm =1-0}, up to $z \lesssim 5$. These SKA capabilities will help fill a critical observational gap at radio wavelengths, offering complementary molecular gas diagnostics to those provided by optical/IR and submillimeter facilities. By tracing the cold molecular gas that fuels early star formation, SKA will help to build a more complete picture of radio galaxy evolution at higher redshifts. In synergy with JWST, ALMA, and future observatories like the ELT and Euclid, SKA will play a key role in advancing our multi-wavelength understanding of galaxy assembly, star formation efficiency, and the evolution of the molecular interstellar medium across cosmic time. 

It is noteworthy that the frequency ranges of Band 5b and Band 6 will also be covered by the next generation Very Large Array (ngVLA) in the future, which will provide exceptional sensitivity to detect the ground transitions of the important molecular tracers mentioned in this paper. However, developing high frequency bands on the SKA still has the advantage to carry out molecular line observations in the southern hemisphere, targeting interesting populations/fields from previous southern sky surveys \citep[e.g., the sources discovered by SPT and ACT,][]{everett20,archipley24, naess25}, or synergic observations/followups with the future facilities such as the ELT and the Vera C. Rubin observatory.

\section{Acknowledgements}
R.W. acknowledges support from the National SKA Program of China (No. 2025SKA0130100), the
Ministry of Science and Technology of the People’s Republic of China under grant No. 2022YFA1602902, and 
the National Natural Science Foundation of China (NSFC) under grant Nos. 12173002, 11991052. 

\bibliographystyle{abbrvnat-maxbibnames4}
\bibliography{chapter} 

\end{document}